\documentclass[11pt]{article}

\begin{document}

\title{Invariant definition of rest mass and dynamics of particles in $4D$ from bulk geodesics in brane-world and non-compact Kaluza-Klein theories}
\author{J. Ponce de Leon\thanks{E-mail: jponce@upracd.upr.clu.edu or jpdel1@hotmail.com}\\ Laboratory of Theoretical Physics, Department of Physics\\ 
University of Puerto Rico, P.O. Box 23343, San Juan, \\ PR 00931, USA}
\date{December 2002}

\maketitle
\begin{abstract}
In the Randall-Sundrum brane-world scenario and other non-compact Kaluza-Klein theories,  the motion of test particles is higher-dimensional in nature. In other words, all test particles travel on five-dimensional geodesics but observers, who are bounded to spacetime, have access only to the $4D$ part of the trajectory. Conventionally, the dynamics of test particles as observed in $4D$ is discussed on the basis of the splitting of the geodesic equation in $5D$. However, this procedure is {\em not} unique and therefore leads to some problems. The most serious one is the ambiguity in the definition of rest mass in $4D$, which is crucial for the discussion of the dynamics. 
We propose the Hamilton-Jacobi formalism, instead of the geodesic one, to study the dynamics in $4D$. On the basis of this formalism we provide an unambiguous expression for the rest mass and its variation along the motion as observed in $4D$. It is independent of the coordinates and any parameterization used along the motion. Also, we are able to show a comprehensive  picture of the various physical scenarios allowed in $4D$, without having to deal with the subtle details of the splitting formalism.
Moreover we  study the extra non-gravitational forces perceived by an observer in $4D$ who describes the geodesic motion of a bulk test particle in $5D$. Firstly, we show that the so-called fifth force fails to account for the variation of rest mass along the particle's worldline. Secondly, we offer here a new definition that correctly takes into account the change of mass observed in $4D$.  

 \end{abstract}



\newpage

\section{Introduction} 

The possibility that our world may be embedded in a $(4 + d)$-dimensional universe with more than four large dimensions has attracted the attention of a great number of researchers.

In higher-dimensional theories of gravity with large extra dimensions, the cylinder condition of the old Kaluza-Klein theory is replaced by the conjecture that the  ordinary matter and fields are confined to a four-dimensional subspace usually referred to as ``3-brane" \cite{Arkani1}-\cite{Arkani3}.  In these theories gravity is a multidimensional interaction that propagates in all dimensions. Randall and Sundrum showed, for $d = 1$, that there is no contradiction between Newton's $1/r^2$ law of gravity in $4D$ and the existence of more that four non-compact dimensions if the background metric is nonfactorizable \cite{RS1}-\cite{RS2}. This has motivated a great deal of active interest in brane-world models where our $4D$ universe is embedded in a five-dimensional non-compact space \cite{Shiromizu}-\cite{Deruelle and Katz}. 

Another non-compact theory of our universe is the so called space-time-matter (STM) theory. In STM the conjecture is that the ordinary matter and fields that we observe in $4D$ result from the geometry of the extra dimension \cite{Wesson 1}-\cite{JPdeL Wesson}

Although brane theory and STM have different physical motivations for the introduction of a large extra dimension, they share the same working scenario, lead to   the same dynamics in $4D$,  and face the same challenges \cite{JPdeLequivalence}. Among them, to predict observationally testable effects of the extra dimension. 
 
The nontrivial dependence of the spacetime metric on the extra coordinate, allowed in both brane-world and STM, implies that the motion of test particles is higher-dimensional in nature. In other words, all particles travel on five-dimensional geodesics but observers, who are bounded to spacetime, have access only to the $4D$ part of the trajectory.

The question of how such an observer, who is unable to access the bulk, perceives the higher-dimensional motion of a test particle has widely been discussed in the literature \cite{MashhoonWesson}-\cite{Seahra2}. The discussion is typically based on the dimensional reduction of geodesics in $5D$. This approach puts forward the possibility of interesting consequences. 

(i) Massless particles in $5D$, with not trivial motion along the extra dimension, are observed as massive particles in $4D$. 

(ii) The rest masses of particles vary as they travel on $5D$ geodesics.

(iii)  There is an anomalous acceleration in $4D$ due to the fifth dimension, or equivalently a ``fifth"  force. 

However, the geodesic approach presents a number of drawbacks. The most important one is that the mass of the particle appears nowhere in the geodesic equation. In this picture, the rest mass in $4D$ is a result of the dimensional reduction of the $5D$ geodesic. The other serious drawback is the ambiguity associated with the choice of affine parameters used to describe the motion in $4D$ and $5D$. Indeed, this ambiguity results in multiple expressions for the particle mass, which in general vary along the particle's worldline. This point was firmly established by Seahra and Wesson in Ref. \cite{Seahra4}  

It is clear that the lack of a proper and unambiguous definition for the rest mass, as  measured in $4D$, crucially affects the applicability, and credibility, of the statements (i)-(iii) mentioned above. 
The object of this paper is to remedy this situation. Our first goal is, therefore, to obtain a definition of rest mass in $4D$ which is independent of the choice of coordinates and geodesic parameters. 

In this work we provide a unified approach for the discussion of the dynamics of test particles in five-dimensional manifolds like those used in brane-world, STM and other non-compact theories. We will discuss (i) the effective rest mass, (ii) the trajectories, and (iii) the forces acting on test particles as observed in $4D$.

In section 2, we present the  invariant definition for the rest mass measured in $4D$. It is based on the invariance of the square of the four-momentum and the Hamilton-Jacobi formalism of Classical Mechanics. This formalism allows us to produce a practical equation for the computation of rest mass, as observed in $4D$, which is completely free of the ambiguities associated with the parameters used in the geodesic approach. This is consistent with the requirement that observables in $4D$ should be invariant under arbitrary transformations in $5D$ \cite{EMT}. This is important for the experimental or observational verification of the theory.

We combine results from the geodesic and the Hamilton-Jacobi methods, in order to provide the general equations for the variation of rest mass as observed in $4D$. We also address the question of under what circumstances a massless $5D$ particle appears as massive in $4D$.  Our approach allows us to give an unambiguous answer  to this question.

Another great advantage of the Hamilton-Jacobi approach is that this equation is a ``scalar" one. Thus, we do not have to deal with the subtle details of dimensional reduction in order to get the adequate  four-dimensional interpretation of the motion in $5D$. In section 3, we discuss this and, with the purpose of illustration of the various physical options in $4D$, we examine the motion of test particles in a specific cosmological setting.

The geodesic motion in the five-dimensional manifold is observed in $4D$ to be under the influence of an extra non-gravitational force.  There are two different versions for this force in the literature. The first one is sometimes called ``fifth force". It is given in terms of a four-dimensional quantity which presents properties atypical of four-vectors \cite{Wesson book}, \cite{MashhoonWesson}-\cite{Youm2}. As a matter of fact, it is not a four-vector. This has been shown by the present author \cite{Eq. of Motion} and confirmed independently by Seahra \cite{Seahra3}  using another approach.  

The second version for this force was presented by this author and exhibits  the appropriate vectorial behavior, but it does not take into account the variation of rest masses \cite{JPdeLforce}, \cite{Eq. of Motion}.  

In section 4, we show another unexpected property of the fifth force. Namely, that it is not related to the change of rest mass in $4D$.  We  show this by using a simple example where the rest  mass is constant and yet the fifth force  is not vanishing. Again, this drawback is a result of the lack of a proper definition for rest mass in $4D$. 

Here we extend our former definition for the extra non-gravitational force to the case of variable rest mass. Our advantage now is that we have  explicit formulae for the variation of mass, they are presented in section $2$. We combine this with some other results to construct an expression for the extra force whose component parallel to $u^{\mu}$ is totally attributable to  the change of rest mass in $4D$.   

\newpage

 \section{The rest mass observed in $4D$} 

The object of this section is to present an invariant definition for the rest mass as measured in $4D$. Our approach is based, not on the geodesic equation, but on the invariance of the square of the four-momentum and it is equivalent to the Hamilton-Jacobi formalism of  Classical Mechanics.

Firstly, we discuss the  attempts to relate  the rest mass of particles to the extra coordinate. Secondly, we use this interpretation as a guide to single out the four-momentum in spacetime from the momentum in $5D$. Finally, we show how the rest mass in $4D$ varies along the observed trajectory in spacetime.

\subsection{Geometrical description of rest mass in $4D$}

There have been a number of attempts to interpret the rest  mass, of uncharged test particles in $4D$,  in  terms of the extra coordinate in $5D$ \cite{Wesson book}. 

The justification for this interpretation comes from the geodesic motion in the five-dimensional background described by the metric
\begin{equation}
\label{canonical metric}
d{\cal{S}}^2 = \frac{y^2}{L^2}g_{\mu\nu}(x^{\rho}, y)dx^{\mu}dx^{\nu} - dy^2,  
\end{equation} 
where $g_{\mu\nu}$ is interpreted as the metric of the spacetime, and $y$ is the ``extra" coordinate. The $0$-component of the five-dimensional geodesic equation gives
\begin{equation}
\label{0-component}
\frac{d}{d{\cal{S}}}\left(\frac{y^2}{L^2}g_{0 \mu}\frac{dx^{\mu}}{d{\cal{S}}}\right)= \frac{y^2}{2 L^2}\frac{\partial{g_{\mu\nu}}}{\partial t}\frac{d x^{\mu}}{d{\cal{S}}}\frac{dx^{\nu}}{d{\cal{S}}}.
\end{equation}
For the case of a static spacetime, $(\partial g_{\mu\nu}/\partial t) = 0$, the quantity in parenthesis is a constant of motion. Besides, one can always choose coordinates such that   $g_{0 j} = 0$ because a static  spacetime should be invariant under the transformation $t \rightarrow - t$. Thus (\ref{0-component}) reduces to
\begin{equation}
\frac{(y/L) \sqrt{g_{00}}}{\sqrt{1 - v^2} \sqrt{1 - (L dy/ y ds)^2}} = C = constant.
\end{equation} 
For geodesic motion with $dy/ds = 0$, the above constant quantity coincides with the energy ${\cal{E}} = m_{o}\sqrt{g_{00}}/\sqrt{1 - v^2}$, provided  we set
\begin{equation}
\label{relation mass and y}
m_{0} = \left(\frac{\cal{E}}{LC}\right)y.
\end{equation}
This expression constitutes the basis for interpreting the  extra coordinate as the $4D$ rest mass. However, it crucially relies upon the  assumptions made on the five-dimensional metric, and the character of the motion in $5D$. In particular that the motion is confined to the hypersurface $y = const$ (or $dy/ds = 0$). Therefore, the only sound conclusion that  seems to follow from (\ref{relation mass and y})  is that, such a motion in  $5D$ is observed in $4D$, as a particle of constant rest mass $m_{0} = ({\cal{E}}/LC)y$.

Although the direct identification of the rest mass with the coordinate $y$ is dubious, for the case under consideration,  the equation (\ref{relation mass and y}) is mathematically correct. We will use it below to get the appropriate connection between the $4D$ part of the five-dimensional momentum and the momentum measured by an observer in $4D$. 

\subsection{Invariant description of rest mass in $4D$}
We now proceed  to discuss the concept of rest mass in $4D$, not in terms of the properties of particular systems of coordinates, but in terms of the invariant definition provided by the normalization of the four-momentum.

In what follows we will consider the background $5D$ metric 
\begin{eqnarray}
\label{general metric}
d{\cal{S}}^2 &=& \Omega(y)g_{\mu\nu}(x^{\rho}, y)dx^{\mu}dx^{\nu} + \epsilon \Phi^2(x^{\rho}, y)dy^2,\nonumber \\
&=& \Omega(y) ds^2 + \epsilon \Phi^2(x^{\rho}, y) dy^2.  
\end{eqnarray}
Where $\Omega(y)$ is called ``warp" factor and satisfies the obvious condition that  $\Omega > 0$, and the factor $\epsilon$ can be $- 1$ or $+ 1$ depending on whether the extra dimension is spacelike or timelike, respectively.  The line element (\ref{general metric}) is more general than  the canonical metric (\ref{canonical metric}) and encompasses all the metrics generally used in brane-world and STM theories.
 
Some  specifics technical details of the discussion, but not the invariant character of $m_{0}$, depend on whether the test particle in $5D$ is massive or massless. We  therefore approach these two cases separately.

\subsubsection{Effective rest mass in $4D$ of {\em massive} particles in $5D$ }
Let us consider a massive test particle moving in the five-dimensional metric (\ref{general metric}).
The momentum $P^A$ of such a particle is defined in the usual way, namely,
\begin{equation}
\label{5D Momentum}
P^{A} = M_{(5)}\left(\frac{dx^{\mu}}{d{\cal{S}}}, \frac{dy}{d{\cal{S}}}\right), 
\end{equation}
where $M_{(5)}$ is the constant five-dimensional mass of the particle and $U^A = (dx^{\mu}/d{\cal{S}}, dy/d{\cal{S}})$ is the velocity in $5D$. Thus $U^AU_A = 1$ and 
\begin{equation}
\label{5D mass}
P^{A}P_{A} = M_{(5)}^2, 
\end{equation}
where the five-dimensional index is lowered and raised with the $5D$ metric. 

The five-dimensional motion is perceived by an observer in $4D$ as the motion of a particle with four-momentum $p_{\mu}$. Consequently,  the  effective rest mass in $4D$ is given by
\begin{equation}
\label{4D mass}
p_{\alpha}p^{\alpha} = m_{0}^2,
\end{equation}
here the four-dimensional indexes are lowered and raised by the spacetime metric $g_{\mu\nu}$.
Because of the absence of cross terms in (\ref{general metric}), the $4D$ components  of $P_{A}$ and $P^{A}$ (i.e., $A = 0, 1,2,3$) are already ``projected" onto spacetime. 

The question is how to identify $P_{\mu}$ or $P^{\mu}$ with the four-momentum $p_{\mu}$ or $p^{\mu}$ in $4D$. 
For  $\Omega = 1$, there is only one possibility, namely $p_{\mu} = P_{\mu}$, $p^{\mu} = P^{\mu}$. Otherwise there are two alternatives;  either (i) $p_{\mu} = P_{\mu}$, $p^{\mu} = \Omega P^{\mu}$, or (ii) $p^{\mu} = P^{\mu}$, $p_{\mu} = \Omega^{- 1} P_{\mu}$. 

For the first alternative, from (\ref{general metric}), (\ref{5D mass}) and (\ref{4D mass}), we get\footnote{Had we assumed $p_{\mu} = \sqrt{W(y)}P_{\mu}$, instead of $p_{\mu} = P_{\mu}$, we would have to replace $\Omega \rightarrow \tilde{\Omega} = \Omega W$ everywhere. This is therefore equivalent to introduce another warp factor, without fundamental changes. }
\begin{equation}
\label{rel between m, P4 and M}
m_{0}^2 + \Omega(y)P_{4}P^{4} = \Omega(y)M_{(5)}^2.
\end{equation}
If the five-dimensional motion is confined to $y = y_{0} = constant$, then $P^4 = 0$ and
\begin{equation}
\label{relation mass and Omega}
m_{0} = \sqrt{\Omega(y_{0})}M_{(5)}.
\end{equation}
Thus, for $\Omega = y^2/L^2 $  we recover (\ref{relation mass and y}) with $M_{(5)} = {\cal{E}}/C$. 

The second alternative, in the case of $P_{4} = 0$ and $\Omega = y^2/L^2$, leads to  a  ``wrong" relation  between $m_{0}$ and $y$, namely $m_{0} \sim y^{- 1}$. Consequently, the correct identification is $p_{\mu} = P_{\mu}$. 

Now, in order to obtain equations of practical use from the above formulae, let us introduce the action $S$ as a function of coordinates, viz., $S = S(x^{\mu}, y)$.  Then, substituting $ - \partial S/ \partial x^{\mu}$ for $P_{\mu}$ and $- \partial S/\partial y$ for $P_{4}$ in (\ref{5D mass}), we obtain the Hamilton-Jacobi equation for a test 
particle in $5D$ 
\begin{equation}
\label{HJ 5D}
g^{\mu \nu}\left(\frac{\partial S}{\partial x^{\mu}}\right)\left(\frac{\partial S}{\partial x^{\nu}}\right) + \frac{\epsilon \Omega}{\Phi^2}\left(\frac{\partial S}{\partial y}\right)^2 = \Omega M^2_{(5)}.
\end{equation}
If we solve this equation, then by virtue of the identification $p_{\alpha} = P_{\alpha}$ and (\ref{4D mass}) we are able to compute  the rest mass measured in $4D$ as
\begin{equation}
\label{HJ 4D}
g^{\mu \nu}\left(\frac{\partial S}{\partial x^{\mu}}\right)\left(\frac{\partial S}{\partial x^{\nu}}\right) = m_{0}^2,
\end{equation}
which is the Hamilton-Jacobi equation for the motion in $4D$. Also, following standard procedure we find the trajectory in $5D$ and the one perceived by an observer in $4D$. 

In the above scheme the action $S$ is a truly higher-dimensional quantity. It carries  all the details of the motion in $5D$ and (\ref{HJ 4D}) acts as  the  ``device"  that  transports the information from $5D$ to $4D$ in an invariant  way. In the next section we provide explicit illustration of this procedure.

The relation between the rest mass in $4D$ and $5D$ is given by
\begin{equation}
\label{relation between the rest mass in 4D and 5D}
m_{0} = \sqrt{\Omega}M_{(5)}\left[1 + \frac{\epsilon \Phi^2}{\Omega}\left(\frac{dy}{ds}\right)^2\right]^{- 1/2}.
\end{equation}
This equation shows how the motion along $y$ affects the rest mass measured in $4D$. It is the five-dimensional counterpart to  $m = m_{0}[1 - v^2]^{-1/2}$, for the variation of particle's mass due to its motion in spacetime. The behavior of $m_{0}$ depends on the signature of the extra dimension. 

\paragraph{Timelike extra dimension:} For $\epsilon = + 1$, the observed $4D$ rest mass decreases as a consequence of motion along $y$. Therefore, it cannot take arbitrary large values, i.e.,
\begin{equation}
 0 < m_{0} \leq \sqrt{\Omega} M_{(5)}.
\end{equation}

\paragraph{Spacelike extra dimension:} For $\epsilon = -1$, it is the opposite and 
\begin{equation}
 \sqrt{\Omega} M_{(5)}\leq m_{0} < \infty.
\end{equation}
 Also, a timelike extra dimension puts no restriction on $(dy/ds)$, while for a spacelike $|dy/ds| < \sqrt{\Omega}/|\Phi|$. 

Let us now use the geodesic equation in $5D$. The $4$-component of this equation gives
\begin{equation}
\label{4-component geod}
\frac{d U_{4}}{d{\cal{S}}} = \frac{1}{2}\frac{\partial (\Omega g_{\mu \nu})}{\partial y} \frac{dx^{\mu}}{d{\cal{S}}}\frac{dx^{\nu}}{d{\cal{S}}}  + \epsilon \Phi \frac{\partial \Phi}{\partial y}\left(\frac{dy}{d{\cal{S}}}\right)^2.
\end{equation}
From this equation, after using (\ref{5D Momentum}) and (\ref{relation between the rest mass in 4D and 5D}), we get  
\begin{equation}
\label{variation of P4}
\frac{1}{m_{0}}\frac{dP_{4}}{ds} = \frac{1}{2\Omega}\frac{\partial(\Omega g_{\mu \nu})}{\partial y} u^{\mu}u^{\nu} +  \frac{\epsilon \Phi}{\Omega}\frac{\partial \Phi}{\partial y}\left(\frac{dy}{ds}\right)^2,
\end{equation}
where $u^{\mu}$ is the usual four-velocity of the particle, viz., $u^{\mu} = (dx^{\mu}/ds)$.
Also, for the variation of the effective rest mass 
\begin{equation}
\label{variation of the effective mass}
\frac{1}{m_{0}}\frac{dm_{0}}{ds} = - \frac{1}{2}u^{\mu}u^{\nu}\frac{\partial g_{\mu\nu}}{\partial y} \frac{dy}{ds} + \frac{\epsilon \Phi u^{\mu}}{\Omega}\frac{\partial \Phi}{\partial x^{\mu}}\left(\frac{dy}{ds}\right)^2.
\end{equation}
The above equations can be expressed, if desired, in terms of the extrinsic curvature of the $y = const.$ hypersurfaces. Indeed, the normal  unit vector ($n_{A}n^{A} = \epsilon$),  orthogonal to these hypersurfaces is given by
\begin{equation}
n^A = \frac{{\delta}_{4}^{A}}{\Phi}, \;\;\;\;\; n_{A} = (0, 0, 0, 0, \epsilon \Phi).
\end{equation}
Thus, the extrinsic curvature $K_{AB} = n_{A;B}$ becomes
\begin{equation}
K_{\mu\nu} = \frac{1}{2 \Phi}\frac{\partial(\Omega g_{\mu\nu})}{\partial y}, \;\;\;\;\; K_{44} = K_{4\mu} = 0.
\end{equation}

\subsubsection{Effective rest mass in $4D$ of {\em massless} particles in $5D$}

Let us now consider a massless test particle moving in the five-dimensional metric (\ref{general metric}). The equation of motion for such a particle is the Eikonal equation, which differs form the one of Hamilton-Jacobi (\ref{HJ 5D}) in that $M_{(5)} = 0$. Also, in (\ref{5D Momentum})  the derivatives $M_{(5)}d/d{\cal{S}}$  have to be replaced by $d/d\lambda$, where $\lambda$ is the parameter along the null geodesic \cite{Landau and Lifshitz}.

The motion of massless particles is along isotropic geodesics, for which $d{\cal{S}} = 0$.  Therefore
\begin{equation}
\label{ds for null geodesics}
\Omega ds^2 = - \epsilon \Phi^2 dy^2.
\end{equation}
Is clear that the signature of the extra dimension plays an important role here.

\paragraph{Timelike extra dimension:}

For $\epsilon = +1$, there is only one physical possibility. From (\ref{relation between the rest mass in 4D and 5D}) it follows that $m_{0} = 0$. That is, massless particles in the so-called two-time $5D$ metrics \cite{Billyard2}-\cite{Kocinski} are perceived as massless in $4D$, they {\em cannot} be observed as massive particles in $4D$. In addition, from (\ref{rel between m, P4 and M}) we get $P_{4} = 0$, i.e., their motion  is confined to hypersurfaces $y = constant$. 

\paragraph{Spacelike extra dimension:}
For $\epsilon = -  1$,  there are two physical possibilities: 

(i) If the massless particle in its motion in $5D$ has non-vanishing velocity along $y$, namely $dy/ds = \sqrt{\Omega}/\Phi$, then it is perceived as a massive particle ($ds^2 > 0$) by an observer in $4D$.

(ii) If the motion of the massless particle in $5D$ is confined to $y = const$, then it is observed as massless ($ds = 0$) particle in $4D$. 

The above discussion can be summarized as follows: null geodesics in $5D$ appear as timelike paths in $4D$ only if the following two conditions are met {\em simultaneously}: the extra dimension is spacelike, and the particle in its five-dimensional motion has $P_{4} \neq 0$. Otherwise, a null geodesic in $5D$ is observed as a lightlike particle in $4D$. 

From (\ref{rel between m, P4 and M}), with $M_{(5)} = 0$, $\epsilon = - 1$ and  $P^{4} = dy/d\lambda$,  we obtain
\begin{equation}
\label{m and P4 for massless particles in 5D}
m_{0} = \sqrt{\Omega} \Phi \frac{dy}{d \lambda} = - \frac{\sqrt{\Omega}}{\Phi}P_{4}.
\end{equation} 
On the other hand, from (\ref{ds for null geodesics}) it follows that $dy =  (\sqrt{\Omega}/\Phi)ds$. Consequently, 
\begin{equation}
\label{geodesic parameter and rest mass}
d \lambda = \left(\frac{\Omega}{m_{0}}\right)ds.
\end{equation}
From this and  the $4$-component of the geodesic equation we obtain
\begin{equation}
\frac{1}{m_{0}}\frac{dP_{4}}{ds} = \frac{1}{2\Omega}\frac{\partial(\Omega g_{\mu \nu})}{\partial y} u^{\mu}u^{\nu} - \frac{1}{\Phi}\frac{\partial \Phi}{\partial y}.
\end{equation}
Now, the variation of rest mass is obtained from (\ref{m and P4 for massless particles in 5D}) as 
\begin{equation}
\label{variation of effective rest mass for massless 5D particles}
\frac{1}{m_{0}}\frac{dm_{0}}{ds} = - \frac{\sqrt{\Omega}}{2 \Phi}\frac{\partial g_{\mu\nu}}{\partial y}u^{\mu}u^{\nu} - \frac{u^{\mu}}{\Phi}\frac{\partial \Phi}{\partial x^{\mu}}. 
\end{equation}
We notice that although the mathematical description of massless particles differs from that of massive particles, the last two equations can be readily obtained from (\ref{variation of P4}) and (\ref{variation of the effective mass}) just by setting $(dy/ds) = (\sqrt{\Omega}/\Phi)$. 

The conclusion from this section is that the effective mass observed in $4D$ can be computed from (\ref{HJ 4D}), where the action is obtained as a solution of (\ref{HJ 5D}) and therefore it transfers  to $4D$ the information of the details of the motion in $5D$.
Another general result is the equation for the variation of rest mass along the trajectory in $4D$, which is given by (\ref{variation of the effective mass}) or (\ref{variation of effective rest mass for massless 5D particles}). 

From a physical point of view, this is important because it shows that the variation of mass $m_{0}$ is {\em not} an artifact of a poor choice of coordinates or parameter in the geodesic equation, but it is an effect from the extra dimension. We also presented a clear four-dimensional interpretation, of the geodesic motion of massless particles in $5D$ for both, spacelike and timelike extra dimension.     

\newpage

\section{The motion observed in $4D$}

The interpretation in $4D$, of the geodesic motion of test particles in $5D$, is usually complicated and obscured by the necessity of working with different affine parameters along the trajectories  in $4D$ and $5D$. As a consequence certain physical quantities seem to depend on the specific choice of parameters. In particular, the changes of $m_{0}$,  may be finite or zero \cite{Seahra4}. 

In this section we present a $4D$ interpretation, of the motion in $5D$, based on the Hamilton-Jacobi equations (\ref{HJ 5D}) and (\ref{HJ 4D}). This interpretation has the advantage of being free of the complications and ambiguities of the geodesic approach. We give an explicit example showing how to obtain the observed rest masses and trajectories in $4D$ from $5D$.

\subsection{Trajectories in $5D$ confined to hypersurfaces $y = constant$}

In this case, from (\ref{relation mass and Omega}) and (\ref{variation of the effective mass}) it follows that $m_{0} = \sqrt{\Omega}M_{(5)}= const.$ along the motion. In particular, a massless particle in $5D$ is observed as a massless particle in $4D$. Notice that (\ref{HJ 5D}) and (\ref{HJ 4D}) become identical to each other for $P_{4} = - ({\partial S}/{\partial y}) = 0$. Thus, the four-dimensional part of the equations of motion in $5D$ is the usual geodesic equation in $4D$.

The question arises of what are the physical requirements to keep the motion of a test particle confined to the hypersurface $y = const.$ 

Several workers have noted that, in general, this confinement requires the introduction of non-gravitational forces. There are a number of methods to derive these forces. Among them, the method of Lagrange undetermined multipliers \cite{Seahra3}, \cite{JPdeLextraforce}. 

However, the answer is provided by our equation  (\ref{variation of P4}). In order to have $dP_{4}/ds = 0$, and thus  prevent the particle from picking pick up momentum along the extra dimension, the non-gravitational  force should be 
\begin{equation}
\frac{f^{(extra)}}{m_{0}} = - \frac{1}{2\Omega}\frac{\partial (\Omega g_{\mu\nu})}{\partial y}u^{\mu}u^{\nu}.
\end{equation}
It is orthogonal to spacetime and therefore not directly measurable by an observer in $4D$. However, Seahra \cite{Seahra3} recently showed that it plays an important role in spinning particles in $5D$.  In a covariant form, this force becomes
\begin{equation}
\label{nongravitational force}
\frac{f^{(extra)}_{A}}{m_{0}} = - \frac{\epsilon}{\Omega}(K_{\mu\nu}u^{\mu}u^{\nu})n_{A}.
\end{equation}
The effect of this non-gravitational force is that, the five-dimensional motion of a particle moving on the hypersurface $y = y_{0}$ is observed in $4D$ as a particle of {\em constant} rest mass (\ref{relation mass and Omega}), moving ``{\em freely}" under the influence of the gravitational field $g_{\mu\nu}$.  This is a general result obtained without imposing any assumptions neither on the spacetime metric $g_{\mu\nu}$ nor on the warp factor $\Omega$.

In the absence of non-gravitational forces, the necessary condition for confinement is 
\begin{equation}
\label{constraint equation}
K_{\mu\nu}u^{\mu}u^{\nu} = 0,
\end{equation}
that is $u^{\mu}u^{\nu}(\partial\Omega g_{\mu\nu}/\partial y) = 0$, which does not imply $K_{\mu\nu} = 0$, in general.
Indeed, (\ref{constraint equation}) is a bilinear combination between the components of the four-velocity, which should be interpreted as a constraint equation. This constraint has to be solved simultaneously with the $4D$ geodesic equation. However, it is not possible to solve the
geodesic equation subject to $K_{\alpha\beta} u^\alpha u^\beta =
0$ in general situations.  Arbitrary hypersurfaces
will have non-trivial curvature that precludes such a
possibility \cite{Seahra5}.  For example, the geodesics of Euclidean 3-space are
straight lines, which cannot be confined to an arbitrarily curved
2-surface like a sphere.  

In higher-dimensional theories the question of which metric frame is the correct representation of our four-dimensional spacetime is a question of practical  importance. In the absence of non-gravitational forces, the condition of confinement requires  test particles to move  on those hypersurfaces where the geodesic equation can be 
solved subject to $K_{\alpha\beta} u^\alpha
u^\beta = 0$. Such hypersurfaces do {\em not} necessarily coincide with the $y = y_{0}$ hypersurfaces. Our conjecture is that the choice of the metric frame in $4D$ should be limited by the condition of  confinement,  without introducing non-gravitational forces.

If neither of the above conditions (\ref{nongravitational force}), (\ref{constraint equation}) are satisfied, then nothing prevents a particle initially having $P_{4} = 0$ from picking  up some momentum along $y$. 

Consider for example a massless particle initially moving in $5D$ with $P_{4} = 0$. 
An observer in $4D$ perceives this motion as a massless $4D$ particle too. However,  
as soon as $P_{4} \neq 0$, the $4D$ rest mass is not longer zero. Indeed, (\ref{rel between m, P4 and M}) 
requires $m_{0} = |P_{4}|\sqrt{\Omega}/\Phi$. 

Thus,  an observer in $4D$ witnesses the  ``spontaneous creation"  of a pair of particles of equal mass and opposite momentum along $y$, viz., $P_{4} = \pm m_{0}\Phi/\sqrt{\Omega}$.

 \subsection{General motion in $5D$: trajectories with $y \neq constant$}

This is the case where the five-dimensional motion has non-vanishing velocity along $y$, i.e., $P_{4} \neq 0$. This is important from an observational viewpoint because, in general, according to (\ref{variation of the effective mass}) the rest masses of  particles measured  by an observer in $4D$, vary along the trajectory. 

The effective rest mass measured in $4D$ is  given by (\ref{HJ 4D}), where the action $S$ is the solution of the five-dimensional Hamilton-Jacobi equation (\ref{HJ 5D}).
Therefore, it implicitly contains the effects due to the large extra dimension and the scalar field. These effects are  explicitly shown in (\ref{variation of the effective mass}) or (\ref{variation of effective rest mass for massless 5D particles}). 

The conclusion from the above discussion is that the observations in $4D$ depend on (i) the mass of the particle in $5D$, (ii) the character of motion in $5D$, and (iii) the nature of the extra coordinate, i.e., whether it is spacelike or timelike. 

\subsection{Trajectories in a cosmological setting}

Here  we examine the motion of a particle in a five-dimensional background which is used to embed standard spatially flat FRW models in $5D$. Our aim  is twofold. Firstly, to show the diversity of scenarios in $5D$ and illustrate in practice how they lead to various physical observations  in $4D$. Secondly, to confirm that the mass observed in $4D$, and indeed the whole description,  is independent and free of the parameters used in the geodesic method.  

\subsubsection{The five-dimensional cosmological  metric} 
 
We  consider the motion of test particles in the background metric \cite{JPdeL 1} 
\begin{equation}
\label{Ponce de Leon solution}
d{\cal S}^2 = y^2 dt^2 - t^{2/\alpha}y^{2/(1 - \alpha)}[dr^2 + r^2(d\theta^2 + \sin^2\theta d\phi^2)] - \alpha^2(1- \alpha)^{-2} t^2 dy^2,
\end{equation}
where $\alpha$ is a constant, $y$ is the coordinate along the  extra-dimension and $t, r, \theta$ and $\phi$ are the usual coordinates for a spacetime with spherically symmetric spatial sections. This is a solution to the five-dimensional Einstein field equations, with ${^{(5)}T}_{AB} = 0$. In what follows we assume that $\Omega = 1$, that is we make no distinction between the induced metric in $4D$ and the metric of the physical spacetime. 

In four-dimensions (on the hypersurfaces $y = const.$) this metric corresponds to the $4D$ Friedmann-Robertson-Walker models with flat $3D$ sections. The  energy density $\rho_{eff}$  and pressure $p_{eff}$ of the effective $4D$ matter satisfy the equation of state
\begin{equation} 
\label{eq of state for the eff fluid}
p_{eff} = n \rho_{eff},  
\end{equation}
where $n = ({2\alpha}/{3} -1)$. Thus  for $\alpha = 2$ we recover radiation, for $\alpha = 3/2$  we recover dust, etc.

 In spherically symmetric fields test particles move on  a single ``plane" passing through the center. We take this plane  as the $\theta = \pi/2$ plane. Then, the Hamilton-Jacobi equation, for the metric (\ref{Ponce de Leon solution}) is 

\begin{equation}
\frac{1}{y^2}\left(\frac{\partial S}{\partial t}\right)^2 - \frac{1}{t^{2/\alpha}y^{2/(1 - \alpha)}}\left[\left(\frac{\partial S}{\partial r}\right)^2 + \frac{1}{r^2}\left(\frac{\partial S}{\partial \phi}\right)^2\right] - \frac{(1 - \alpha)^2}{\alpha^2 t^2}\left(\frac{\partial S}{\partial y}\right)^2 = M_{(5)}^2.
\end{equation}
Since $\phi$ is a cyclic coordinate, it is clear that the action separates as
\begin{equation}
S = S_{1}(t, y) + S_{2}(r) + L \phi,
\end{equation}
where $L$ is the angular momentum.  
Thus, we obtain
\begin{equation}
\label{S1}
\frac{1}{y^2}\left(\frac{\partial S_{1}}{\partial t}\right)^2 - \frac{k^2}{t^{2/\alpha}y^{2/(1 - \alpha)}}  - \frac{(1 - \alpha)^2}{\alpha^2 t^2}\left(\frac{\partial S_{1}}{\partial y}\right)^2 = M_{(5)}^2,
\end{equation}
and
\begin{equation}
\left(\frac{dS_{2}}{dr}\right)^2 + \frac{L^2}{r^2} = k^2 \geq 0,
\end{equation}
where $k$ is the separation constant. If it is zero, then the particle is commoving (or at rest) in the system of reference defined by (\ref{Ponce de Leon solution}).

From (\ref{HJ 4D}) we obtain the rest mass, as measured in $4D$, as follows
\begin{equation}
\label{effective mass as measured in 4D}
m_{0}^2 = \frac{1}{y^2}\left(\frac{\partial S_{1}}{\partial t}\right)^2 - \frac{k^2}{t^{2/\alpha}y^{2/(1 - \alpha)}},
\end{equation}
which has to be evaluated at the trajectory $y = y(t)$.
Thus, for any given solution of  (\ref{S1}) we can calculate  the corresponding rest mass  observed in $4D$. This prescription is totally free of the ambiguities, typical of the approach where the rest mass in $4D$ is obtained from the dimensional reduction of the geodesic equation in $5D$. 

We  now proceed to consider different physical scenarios allowed in (\ref{S1}). It should be noted that the background metric (\ref{Ponce de Leon solution}) has recently been considered in Ref. \cite{JPdeLQM} in the ``classical" context where $P_{4}$ can be identified with the electrical charge.

\subsubsection{Confined motion}
 
In the case where $P_{4} = - (\partial S/\partial y) = 0$, equation (\ref{S1}) requires  $y = const$, and the effective rest mass observed in $4D$ is  $m_{0} = M_{(5)}$, as expected. 

There are two possibilities. Namely, either the particle is at rest ($k = 0$), or it is somehow moving in spacetime ($k \neq 0$). In order to isolate the effects of the extra dimension, from the effects due to the motion in spacetime, we will consider $k = 0$.
Thus, from (\ref{effective mass as measured in 4D}) we get 
\begin{equation}
S_{1} = - m_{0}y_{0}t.
\end{equation} 
Consequently, $P_{0} = -(\partial S/\partial t) = m_{0}y_{0}$ and $P^{0} = m_{0}/y_{0}$. Thus $u^{\mu} = \delta^{\mu}_{0}/y_{0}$, which correctly satisfies $g_{\mu\nu} u^{\mu}u^{\nu} = 1$. In this case, the non-gravitational ``force" needed to keep the particle on the hypersurface $y = y_{0}$ is $f^{(extra)}_{A} = - (m_{0}/y_{0})\delta_{A}^{4}$.

\subsubsection{General motion of massive test particles: $M_{(5)} \neq 0$}

For a particle at rest in spacetime ($k = 0$), from (\ref{S1}) one can easily get
\begin{equation}
S_{1}(t, y) = \pm \frac{\alpha M_{(5)}}{\sqrt{2\alpha - 1}} yt.
\end{equation}
Consequently, 
\begin{equation}
\label{rest mass for massive 5D part in cosmological setting}
m_{0} = \frac{\alpha M_{(5)}}{\sqrt{2\alpha - 1}}, \;\;\;\;\;P_{4} =  \pm m_{0}t.
\end{equation}
Here the rest mass is constant in time, because of the mutual cancelation of the change induced by the term $(\partial g_{\mu\nu}/\partial y)u^{\mu}u^{\nu}$ and the change induced by the scalar field. However it does depend on the epoch, that is on  $\alpha$.
Now, using 
\begin{eqnarray}
\label{Momentum}
P^{\rho} &=& M_{(5)}\frac{dx^{\rho}}{d{\cal{S}}} = g^{\rho \lambda}P_{\lambda} = - g^{\rho \lambda}\frac{\partial S}{\partial x^{\lambda}},\nonumber \\
P^{4} &=& M_{(5)}\frac{dy}{d{\cal{S}}} = \frac{\epsilon}{\Phi^2}P_{4} = - \frac{\epsilon}{\Phi^2}\frac{\partial S}{\partial y}
\end{eqnarray}
we obtain
\begin{equation}
\frac{dt}{d\cal{S}}  = \mp \frac{\alpha}{y \sqrt{2\alpha -1}},
\end{equation}
and
\begin{equation}
\frac{dy}{d\cal{S}} = \pm \frac{(1 - \alpha)^2}{\alpha t\sqrt{2\alpha -1}}.
\end{equation}
From these expressions we evaluate $dy/dt$ and integrate to obtain
\begin{equation}
\label{trajectory}
y = Dt^{- (1 - \alpha)^2/\alpha^2},
\end{equation}
along the trajectory, where $D$ is a constant of integration. From here we obtain $dy/ds$ as follows
\begin{equation}
\label{dy/ds for sol with constant mass}
\frac{dy}{ds} = \frac{1}{y}\frac{dy}{dt} = - \frac{(1 - \alpha)^2}{{\alpha}^2 t}.
\end{equation}
Also we find the four-velocity $u^{\mu} = \delta^{\mu}_{0}/y$ and $u^4 = |(\alpha - 1)/\alpha|$. The proper time $\tau$ along the trajectory is given by $d\tau = y(t)dt$. Thus, from (\ref{trajectory}), we get $\tau \sim t^{(2 \alpha -1)/\alpha^2}$. Consequently, in this case, $m_{0}$ is a constant and $P_{4} \sim \tau^{\alpha^2/(2\alpha -1)}$. 

\subsubsection{General motion of massless test particles: $M_{(5)} = 0$}

In this case the trajectory in $5D$ is along isotropic geodesics. These are given by the Eikonal equation, which is formally obtained from the above formulae by setting $M_{(5)} = 0$ in (\ref{S1}).
According to the above discussion, there are two different physical possibilities here. They are   $P_{4} \neq 0$, or $P_{4} = 0$. 

Firstly we study the case where $M_{(5)} = 0$ and $P_{4} \neq 0$. 
For the same reason as above, we consider $k = 0$. Then, equation (\ref{S1}) separates and, we obtain
\begin{equation}
S_{1} = C t^{\pm l}y^{\pm l \alpha/(1 - \alpha)},
\end{equation} 
where $C$ is a constant of integration and $l$ is the separation constant. Using (\ref{HJ 4D}) we find the rest mass 
\begin{equation}
\label{rest mass as a function of t and y}
m_{0} = |l||C|t^{- 1 \pm l}y ^{- 1 \pm l \alpha/(1 - \alpha)},
\end{equation}
which has to be evaluated along the trajectory. 
Now in (\ref{Momentum}) instead of the derivatives $M_{(5)}d/d{\cal{S}}$ we have to write derivatives $d/d\lambda$, where $\lambda$ is the parameter along the null geodesic. Thus, we find
\begin{equation}
\label{t as a function of lambda}
\frac{dt}{d\lambda} = \mp C l t^{(- 1 \pm l)}y^{(- 2 \pm l \alpha/(1 - \alpha))},
\end{equation}
and
\begin{equation}
\label{y as a function of lambda}
\frac{dy}{d\lambda} = \pm C l \frac{1 - \alpha}{\alpha}t^{(- 2 \pm l)} y ^{(- 1 \pm l \alpha/(1 - \alpha))}.
\end{equation}
Since $dt/ds = 1/y$, from (\ref{geodesic parameter and rest mass}), (\ref{rest mass as a function of t and y}) and (\ref{t as a function of lambda}) it follows that
\begin{equation}
\mp \frac{C l}{|C||l|} = + 1.
\end{equation}
Consequently, from (\ref{y as a function of lambda}) 
\begin{equation}
\label{y as a function of s}
\frac{dy}{ds} = \frac{1}{\Phi} = - \frac{(1 - \alpha)}{\alpha t},
\end{equation}
as expected. From the above expressions we evaluate $dy/dt$ and integrate to obtain $y = Dt^{(\alpha - 1)/\alpha}$.  Consequently, in this case we find $P_{4}$ and $m_{0}$ as follows
\begin{eqnarray}
P_{4} &=& A \alpha (1 - \alpha)^{- 1}t^{(1 - \alpha)/\alpha},\nonumber \\
m_{0} &=& A t^{(1 - 2\alpha)/\alpha},
\end{eqnarray}
where $A$ is expressed through the other constants as $A = |C||l| D^{(- 1 \pm l \alpha /(1 - \alpha))}$. Notice that, (\ref{m and P4 for massless particles in 5D}) is identically satisfied by virtue of (\ref{y as a function of s}). 

For the proper time we find $\tau = (D \alpha/(2\alpha -1)) t^{(2\alpha - 1)/\alpha}$. Thus, here $P_{4} \sim \tau^{(1 - \alpha)/(2\alpha - 1)}$ and $m_{0} \sim \tau^{- 1}$. Again, we notice the advantage of our approach here, namely that neither $m_{0}$ nor $P_{4}$ (nor their variations) depend on the choice of parameters used in the geodesic description of the $5D$ or $4D$ motion, as one expected.

\medskip

Secondly we study the case where $M_{(5)} = 0$ and $P_{4} = 0$.
In this case the motion, as observed in $4D$, is lightlike and, therefore, $k$ must be different from zero. The left hand side in (\ref{effective mass as measured in 4D})  is $m_{0} = 0$. This equation is equivalent to $k^{\mu}k_{\mu} = 0$, where $k_{\mu}$ is the $4D$ wave vector.  Since $P_{4} = 0$, it follows that $y = const$ along the motion. Therefore, the frequency $\omega = - \partial S/ \partial t$ of the ``induced" photon is 
\begin{equation}
\omega \sim \tau^{- 1/\alpha}.
\end{equation} 

\subsubsection{Spacelike ($d{\cal{S}}^2 < 0$) trajectories in $5D$ }

Finally, for completeness we mention that for a spacelike extra dimension ($\epsilon = -1 $) there is one more possibility left here. Namely the motion in $5D$ with $d{\cal{S}}^2 < 0$. In this case replacing $M_{(5)}^2 \rightarrow - {\bar{M}}^2$ we obtain the same expressions as in section $3.3.3$, but with $\bar{M}$ instead of $M_{(5)}$ and $\sqrt{1 - 2\alpha}$ instead of $\sqrt{2\alpha - 1}$. Thus,  a five-dimensional spacelike interval ($d{\cal{S}}^2 < 0$)  can be  interpreted by an observer in $4D$ as a test particle with positive effective rest mass ($ds^2 > 0$). This is a pure consequence of the motion in $5D$ along the {\em spacelike} extra coordinate. We also notice that the metric (\ref{Ponce de Leon solution}) with $\alpha < 1/2$ describes inflationary models that expand faster than the standard FRW ones.

\medskip

The conclusion  from this section is that the Hamilton-Jacobi formalism allows us to give a complete description of the motion in $5D$ and $4D$, including the explicit evaluation of the effective rest mass and trajectories in $4D$, without having to deal with the details, and ambiguities, associated with the splitting of the five-dimensional geodesic equation.  

\medskip

From  a physical point of view,  our model exhibits an interesting behavior. Namely that  $dy/ds \sim t^{-1}$ for both scenarios in $3.3.3$ and $3.3.4$ (equations (\ref{dy/ds for sol with constant mass}) and (\ref{y as a function of s})). Which means that, although particles move in the extra direction, their motion becomes confined to some $y = y_{0}$ hypersurface asymptotically as $t\rightarrow \infty$. On such hypersurface the metric (\ref{Ponce de Leon solution}) corresponds to the $4D$ FRW models with flat $3D$ sections. Therefore, this is a good physical property,   consistent with the paradigm that the matter is confined to our four-dimensional spacetime.  
\newpage
\section{Forces in brane-world, STM and other non-compact $5D$ theories}

So far we have learned how to get the effective mass and trajectories in $4D$ from the geodesic motion in $5D$. We now turn our attention to the dynamics observed in $4D$. It is clear that the bulk geodesic motion of a test particle is observed in $4D$ as the motion of a particle under the influence of a non-gravitational force. 

In special relativity the four-force acting on a test particle is defined by
\begin{equation}
\label{force in special relativity}
F^{\mu} = \frac{d p^{\mu}}{ds} = \frac{d}{ds}(m_{0}u^{\mu}).
\end{equation}
In curvilinear coordinates the generalization of (\ref{force in special relativity}), in contravariant notation, is 
\begin{equation}
\label{Definition of force in 4D contravariant}
F^{\mu} = \frac{Dp^{\mu}}{ds} = \frac{d p^{\mu}}{ds} + \Gamma^{\mu}_{\alpha \beta}u^{\alpha}p^{\beta},
\end{equation}
and in covariant notation it is
\begin{equation}
\label{definition of force in 4D covariant}
F_{\mu} = \frac{Dp_{\mu}}{ds} = \frac{d p_{\mu}}{ds} - \frac{1}{2}\frac{\partial g_{\alpha\beta}}{\partial x^{\mu}}u^{\alpha}p^{\beta}.
\end{equation}
Since $u^{\mu}u_{\mu} = 1$, it follows that the acceleration $a^{\mu} = Du^{\mu}/ds = g^{\mu \nu} Du_{\nu}/ds$ is orthogonal to the four-velocity. If the  rest mass is constant, then $F^{\mu}u_{\mu} = 0$. Therefore, the variation of rest mass is observed as a force acting parallel to the particles's four-velocity, namely,
\begin{equation}
\label{parallel component}
F^{\mu}_{\parallel} = u^{\mu}\frac{dm_{0}}{ds}
\end{equation}  
In this section we consider the same scenario as in sections $2$ and $3$.  
Namely, the geodesic motion of test particles on $5D$ manifolds like those used in brane-world, STM and other non-compact $5D$ theories. Our aim is to provide the appropriate expression for the non-gravitational force perceived by an observer in $4D$. 

Our advantage here, with respect to previous work on the subject, is that we have an explicit invariant equation, namely (\ref{variation of the effective mass}) or (\ref{variation of effective rest mass for massless 5D particles}), for the variation of rest mass along the particle's worldline. 

First, we demonstrate the flaws of (\ref{Definition of force in 4D contravariant}) and (\ref{definition of force in 4D covariant}) when applied to non-compact Kaluza-Klein theories. Second, we offer a new definition of force which exhibits good physical properties and appropriately takes into account the variation of rest mass. 

\subsection{Testing the definitions}
Let us start our discussion by testing how well or how bad  the ``classical"  definitions (\ref{Definition of force in 4D contravariant}) and (\ref{definition of force in 4D covariant}) work when applied to some particular five-dimensional motion.

The simplest test is to apply the definitions in the scenario where the  particle moving in  $5D$ is perceived by an observed in $4D$ as a particle of {\em constant} rest mass $m_{0}$. 

Such a scenario  is provided by the solution discussed in section $3.3.3$, for which  $m_{0} = \alpha M_{(5)}/\sqrt{2\alpha - 1}$. Substituting $u^{\mu} = \delta^{\mu}_{0}/ y$, and using (\ref{Ponce de Leon solution}), into  (\ref{Definition of force in 4D contravariant}) we easily get
\begin{equation}
\label{fifth force contr}
\frac{1}{m_{0}}{F^{\mu}_{\parallel}} = \frac{1}{m_{0}}\frac{Dp^{\mu}}{ds} = - \frac{\delta^{\mu}_{0}}{y^2}\frac{dy}{ds} = - \frac{u^{\mu}}{y}\frac{dy}{ds}.
\end{equation}
This is an unexpected result, because $m_{0}$ is constant here and we should have obtained $F^{\mu}_{\parallel} = 0$ instead, as predicted by (\ref{parallel component}). For the case under consideration, the right hand side of (\ref{fifth force contr}) can be written as
\begin{equation}
\label{fifth force contr General expression}
\frac{1}{m_{0}} F^{\mu}_{\parallel} = - \frac{u^{\mu}}{2}\frac{\partial g_{\alpha\beta}}{\partial y}u^{\alpha}u^{\beta}\frac{dy}{ds},
\end{equation}
which exactly coincides with the so-called fifth-force\footnote{See for example page $166$ in Ref \cite{Wesson book}.}.

If we now substitute $u_{\mu} = y \delta_{\mu}^{0}$ into (\ref{definition of force in 4D covariant}) we obtain
\begin{equation}
\label{fifth force cov} 
\frac{1}{m_{0}} F_{\parallel \mu} = \frac{1}{m_{0}}\frac{Dp_{\mu}}{ds}  = \delta_{\mu}^{0}\frac{dy}{ds} = \frac{u_{\mu}}{y}\frac{dy}{ds}.
\end{equation}
The non-vectorial properties of $F^{\mu}_{\parallel}$ and/or $F_{\parallel \mu}$, given by  (\ref{fifth force contr})-(\ref{fifth force cov}), are immediately obvious. Firstly,  
\begin{equation}
\label{prop 1}
u_{\mu}F^{\mu}_{\parallel} \neq u^{\mu}F_{\parallel \mu}.
\end{equation}
Secondly,
\begin{equation}
\label{prop 2}
F_{\parallel\mu} \neq g_{\mu\nu}F^{\nu}_{\parallel}.
\end{equation}

In conclusion, (\ref{Definition of force in 4D contravariant}) and (\ref{definition of force in 4D covariant}) lead to a force, parallel to $u^{\mu}$, with two surprising properties that we regard as deficiencies. 

The first one is that $F^{\alpha}_{\parallel}$ is {\em not} a  four-vector. This has recently been discussed, for constant rest mass, in the literature \cite{JPdeLforce}, \cite{Seahra3}, \cite{Eq. of Motion}. 

The second property, is that (\ref{fifth force contr General expression}) is {\em not} related to the change of rest mass, because we anticipated $F^{\alpha}_{\parallel} = 0$ for $m_{0} = const$.

\subsection{Properties of $F^{\sigma} = Dp^{\sigma}/ds$ and $F_{\sigma} = Dp_{\sigma}/ds$}
We now proceed to show that the abnormal properties of $F^{\mu}_{\parallel}$ and $F_{\parallel \mu}$, are not particular to the model discussed above, but are generic properties of $Dp^{\sigma}/ds$ and $Dp_{\sigma}/ds$.  

The usual approach  for obtaining the equations for the four velocity of a test particle is to decompose the five-dimensional geodesic equation into a $4D$ part and a part governing the motion in the extra dimension. Then, the set of equations in $4D$ is manipulated to isolate an expression for $Du^{\mu}/ds$. 

In what follows we set $\Omega = 1$. This simplifies  the notation and does not affect the generality of the discussion. 

\subsubsection{Computing $F^{\sigma} = Dp^{\sigma}/ds$}
For the line element (\ref{general metric}) we obtain \cite{JPdeLforce}, \cite{Eq. of Motion}
\begin{equation}
\label{4D part in noncompact KK}
\frac{Du^{\sigma}}{ds} =  
\epsilon \Phi (\frac{dy}{ds})^2\left[\Phi^{;\sigma} - u^{\sigma}\Phi_{;\rho}u^{\rho} \right] + \left(\frac{1}{2}u^{\sigma}u^{\lambda} - g^{\sigma\lambda}\right)u^{\rho} \frac{\partial {g_{\lambda \rho}}}{\partial y}\frac{dy}{ds},
\end{equation}
where $\Phi_{; \alpha} = \bigtriangledown_{\alpha}\Phi$. Now using (\ref{variation of the effective mass}) and 
\begin{equation}
F^{\alpha} = \frac{Dp^{\alpha}}{ds} = m_{0}\frac{Du^{\alpha}}{ds} + u^{\alpha}\frac{dm_{0}}{ds},
\end{equation}
we obtain
\begin{equation}
\label{Classical force contr per unit mass}
 \frac{1}{m_{0}}\frac{Dp^{\sigma}}{ds} = - g^{\sigma \lambda} u^{\rho}\frac{\partial g_{\lambda \rho}}{\partial y}\frac{dy}{ds} + \epsilon \Phi \Phi^{; \sigma}\left(\frac{dy}{ds}\right)^2.
\end{equation}
If we set $\Phi = constant$, this expression reduces to equation $(55)$ in Ref. \cite{Youm1} with $\Omega = 1$. The force (\ref{Classical force contr per unit mass}) has a component parallel to $u^{\sigma}$ which is given by\footnote{The parallel component of $(55)$ in Ref. \cite{Youm1} differs from the one in Ref. \cite{Wesson book} by a factor $ 1/2$.}  
\begin{equation}
\label{parallel component of general class. def. of force}
\frac{1}{m_{0}}F^{\sigma}_{\parallel} = - \frac{u^{\sigma}}{2}\frac{\partial g_{\alpha \beta}}{\partial y}u^{\alpha}u^{\beta} \frac{dy}{ds} + \frac{u^{\sigma}}{m_{0}}\frac{dm_{0}}{ds}.
\end{equation}
Notice the similarity between this equation and (\ref{fifth force contr General expression}). It is clear that the first term, which  is identical to the fifth force in (\ref{fifth force contr General expression}), is present all the time regardless of whether $m_{0}$ is constant or not. In other words the fifth force term in (\ref{fifth force contr General expression}) and (\ref{parallel component of general class. def. of force}) is a geometrical term {\em not} related to the variation of mass. 
  
\subsubsection{Computing $F_{\sigma } = Dp_{\sigma}/ds$}
The same procedure for $u_{\sigma}$ yields
\begin{equation}
\label{f lit cov}
\frac{Du_{\sigma}}{ds} = \epsilon \Phi (\frac{dy}{ds})^2\left[\Phi_{;\sigma} - u_{\sigma}\Phi_{;\rho}u^{\rho} \right] + \frac{1}{2}u_{\sigma}u^{\lambda}u^{\rho}\frac{\partial {g_{\lambda\rho}}}{\partial y}\frac{dy}{ds}.
\end{equation}
Again, using (\ref{variation of the effective mass}) and 
\begin{equation}
F_{\alpha} = \frac{Dp_{\alpha}}{ds} = m_{0}\frac{Du_{\alpha}}{ds} + u_{\alpha}\frac{dm_{0}}{ds},
\end{equation}
we obtain
\begin{equation}
\label{Classical force cov per unit mass}
\frac{1}{m_{0}}\frac{Dp_{\sigma}}{ds} =  \epsilon \Phi \Phi_{; \sigma}\left(\frac{dy}{ds}\right)^2.
\end{equation}
We immediately notice that (\ref{Classical force contr per unit mass}) and (\ref{Classical force cov per unit mass}) appropriately reproduce  the previous expressions (\ref{fifth force contr}) and (\ref{fifth force cov}), for the metric (\ref{Ponce de Leon solution}) and   $dy/ds$ from (\ref{dy/ds for sol with constant mass}).

\medskip

The above expressions show that, in general, $F^{\sigma} = Dp^{\sigma}/ds$ and $F_{\sigma} = Dp_{\sigma}/ds$ satisfy non-vectorial relations similar to those in (\ref{prop 1}) and (\ref{prop 2}). In addition,
\begin{equation}
\label{prop 3}
F_{\alpha} = g_{\alpha \beta}F^{\beta} + p^{\beta}\frac{\partial g_{\alpha \beta}}{\partial y}\frac{dy}{ds}.
\end{equation}
Thus, they behave as contravariant and covariant components of a four-vector only 
in classical Kaluza-Klein theory, where the ``cylinder" condition $(\partial g_{\mu\nu}/\partial y) = 0$  is imposed. This result extends the previous studies  \cite{JPdeLforce}, \cite{Seahra3}, \cite{Eq. of Motion} to include the variation of the rest masses of particles as observed in $4D$.

\bigskip

The new and surprising result here is that the fifth force is not related to the variation of mass because it does not vanish even when the  mass $m_{0}$ is constant.

\subsection{The extra force in  Kaluza-Klein theory  without cylindricity}

This is the scenario in membrane theory and STM, where the five-dimensional metric is allowed to depend explicitly on the extra coordinate. 

What we have seen so far is that the use of definitions (\ref{Definition of force in 4D contravariant}) and (\ref{definition of force in 4D covariant}) in non-compact Kaluza-Klein theory leads to a force whose properties contradict usual physics in $4D$. Some authors argue that these unusual properties, and the corresponding violation of the laws of physics in $4D$, are inevitable consequence  of large extra dimensions \cite{WessonMashhoon}, \cite{Youm1}, \cite{WessonClassQuanGrav}. 

However, this interpretation is not unique, and an alternative point of view is possible. In order to see this, let us examine the roots of these unusual properties. 

Firstly, notice that $Dg_{\mu\nu} \neq 0$. Indeed, 
\begin{equation}
\label{Abs. Dif. of metric in 5D}
Dg_{\mu\nu}= \left[g_{\mu\nu,\rho} - \left({\Gamma}^{\lambda}_{\mu\rho}g_{\lambda\nu}+ {\Gamma}^{\lambda}_{\nu\rho}g_{\lambda\mu}\right) \right]dx^{\rho} + \frac{\partial{g_{\mu\nu}}}{\partial y}dy.
\end{equation}
The term in square bracket is the absolute differential in $4D$, which we will denote as $D^{(4)}$. For which $D^{(4)}g_{\mu\nu} = 0$. Consequently, 
\begin{equation}
Dg_{\mu\nu} = \frac{\partial g_{\mu \nu}}{\partial y}dy,\;\;\;\;\;\;\;\;\;\; Dg^{\mu \nu} = - g^{\mu \lambda}g^{\nu \rho}\frac{\partial g_{\lambda \rho}}{\partial y}dy.
\end{equation}
These expressions serve to prove the consistency between 
 (\ref{fifth force contr}) and (\ref{fifth force cov}), or between (\ref{Classical force contr per unit mass}) and (\ref{Classical force cov per unit mass}).

Secondly, notice that $u_{\mu}Du^{\mu} \neq 0$, $u^{\mu}Du_{\mu} \neq 0$, and $u_{\mu}Du^{\mu} \neq u^{\mu}Du_{\mu}$. In fact, from (\ref{4D part in noncompact KK}) we find 

\begin{equation}
u_{\sigma}\left[\frac{Du^{\sigma}}{ds} - \frac{\partial u^{\sigma}}{\partial y} \frac{dy}{ds}\right] = 0,
\end{equation}
where 
\begin{equation}
\label{calculation of partial derivative of u, part 3}
\frac{\partial{u^{\mu}}}{\partial y}= - \frac{1}{2}u^{\mu}\frac{\partial{g_{\alpha\beta}}}{\partial y}u^{\alpha}u^{\beta},
\end{equation}
which can be easily calculated in the comoving frame where $u^{\mu} = \delta ^{\mu}_{0}/\sqrt{g_{00}}$, or can be obtained from ``first principles" as in Refs. \cite{JPdeLforce} and \cite{Eq. of Motion}. A similar expression can be obtained for (\ref{f lit cov}), viz., 
\begin{equation}
u^{\sigma}\left[\frac{Du_{\sigma}}{ds} - \frac{\partial u_{\sigma}}{\partial y} \frac{dy}{ds}\right]  = 0.
\end{equation}
The above formulae seem to be incompatible with the usual properties of the metric and four-velocity  in general relativity and compactified  Kaluza-Klein theory. However, if we define $D^{(4)}$, the ``operator"  for absolute differential in $4D$, as 
\begin{equation}
\label{operator D4}
D^{(4)} = D - dy \frac{\partial}{\partial y},
\end{equation}
then, the above equations can be cast into a more familiar form. Namely $D^{(4)}g_{\mu\nu} = D^{(4)}g^{\mu \nu} = 0$ and $u_{\mu}D^{(4)}u^{\mu} = u^{\mu}D^{(4)}u_{\mu} = 0$.

What comes out from the above discussion is the suggestion that we can {\em eliminate} the contradictions and unusual properties of the extra force by  making a ``small" reinterpretation of (\ref{Definition of force in 4D contravariant}) and (\ref{definition of force in 4D covariant}). 

\subsubsection{New definition for the extra force}

Thus, the above equations do not contradict physics in $4D$ if we interpret them appropriately. We propose to define the four-acceleration through $D^{(4)}u^{\mu}$ instead of $Du^{\mu}$, namely,
\begin{equation}
\label{correct definition of acceleration}
a^{\mu}= \frac{D^{(4)}u^{\mu}}{ds}, \;\;\;\;\;\  a_{\mu}= \frac{D^{(4)}u_{\mu}}{ds}.
\end{equation}
This is consistent with  $u_{\mu}a^{\mu} = u^{\mu}a_{\mu} = 0$, although $u_{\mu}Du^{\mu} \neq u^{\mu}Du_{\mu} \neq 0$. Also, since $D^{(4)}g_{\mu\nu} = 0$, we have $a_{\sigma} = g_{\sigma\mu}a^{\mu}$, as desired.

Moreover, as a consequence of  (\ref{calculation of partial derivative of u, part 3}), with  this definition we take away the spurious parallel acceleration  (force per unit mass), which is not related to any change of mass in (\ref{fifth force contr General expression}) and the first term in (\ref{parallel component of general class. def. of force}). Namely, 
\begin{equation}
f^{\mu} = F^{\mu} - m_{0}\frac{\partial u^{\mu}}{\partial y}\frac{dy}{ds} = m_{0} \frac{D^{(4)}u^{\mu}}{ds} + u^{\mu}\frac{dm_{0}}{ds}.
\end{equation}
Here and in what follows we use lower case $f$ to denote this ``new" definition of force. The explicit form of this force is as follows
\begin{equation}
\label{new def for force per unit mass contr}
\frac{1}{m_{0}}f^{\mu} = \epsilon \Phi [\Phi^{\mu} - u^{\mu}\Phi_{\rho}u^{\rho}]\left(\frac{dy}{ds}\right)^2 + [u^{\mu}u^{\rho} - g^{\mu\rho}] u^{\lambda}\frac{\partial g_{\rho \lambda}}{\partial y} \frac{dy}{ds} + \frac{u^{\mu}}{m_{0}}\frac{dm_{0}}{ds}.
\end{equation}
For the covariant components we find 
\begin{equation}
\label{new def for force per unit mass cov}
\frac{1}{m_{0}}f_{\mu} = \epsilon \Phi [\Phi_{\mu} - u_{\mu}\Phi_{\rho}u^{\rho}]\left(\frac{dy}{ds}\right)^2 + [u_{\mu}u^{\rho} - \delta^{\rho}_{\mu}] u^{\lambda}\frac{\partial g_{\rho \lambda}}{\partial y} \frac{dy}{ds} + \frac{u_{\mu}}{m_{0}}\frac{dm_{0}}{ds}.
\end{equation}
Notice that in our version the contravariant and covariant components of the force satisfy the usual requirements for four-vectors.  The last term in the above formulae extends our previous results  \cite{JPdeLforce}, \cite{Eq. of Motion} to include the variation of the rest masses\footnote{We note that in  \cite{JPdeLforce} and \cite{Eq. of Motion} the point in discussion is the lack of vectorial properties of (\ref{Definition of force in 4D contravariant}) and (\ref{definition of force in 4D covariant}) when applied to non-compact Kaluza-Klein theories. Therefore the mass $m_{0}$ was taken as constant.} of particles as observed in $4D$. 
 
To summarize, a massive bulk test particle ($M_{(5)} \neq 0$) moving freely in a five-dimensional manifold is observed in $4D$ as a massive particle ($m_{0} \neq 0$)  moving under the influence of the force given by (\ref{new def for force per unit mass contr}) and/or (\ref{new def for force per unit mass cov}). It contains three distinct contributions, viz.,  

\begin{equation}
f^{\mu} = f^{\mu}_{\Phi \perp} + f^{\mu}_{g \perp} + f^{\mu}_{\parallel},
\end{equation}
where
\begin{equation}
\frac{1}{m_{0}}f^{\mu}_{\Phi \perp} = \epsilon \Phi [\Phi^{\mu} - u^{\mu}\Phi_{\rho}u^{\rho}]\left(\frac{dy}{ds}\right)^2,
\end{equation}
\begin{equation} 
\frac{1}{m_{0}}f^{\mu}_{g \perp} = [u^{\mu}u^{\rho} - g^{\mu\rho}] u^{\lambda}\frac{\partial g_{\rho \lambda}}{\partial y} \frac{dy}{ds},
\end{equation}
and from (\ref{variation of the effective mass})
\begin{equation}
\label{extra force parallel to four velocity, new definition}
\frac{1}{m_{0}}f^{\mu}_{\parallel} = u^{\mu}\left[- \frac{1}{2}u^{\lambda}u^{\rho}\frac{\partial g_{\lambda\rho}}{\partial y} \frac{dy}{ds} + {\epsilon \Phi u^{\lambda}}\frac{\partial \Phi}{\partial x^{\lambda}}\left(\frac{dy}{ds}\right)^2\right].
\end{equation}
All these terms have to be evaluated along the trajectory. 

Now, the bulk geodesic motion of a massless  particle ($M_{(5)} = 0$) with $dy/ds \neq 0$ is observed in $4D$ as the motion of a massive particle ($m_{0} \neq 0$) under the influence of the force given by (\ref{new def for force per unit mass contr}) and/or (\ref{new def for force per unit mass cov}) with $dy/ds$ replaced by $1/\Phi$.    

The first two terms $f^{\mu}_{\Phi \perp}$ and $f^{\mu}_{g \perp}$ have already been discussed in \cite{JPdeLforce}, \cite{Eq. of Motion}. They are orthogonal to the four-velocity and therefore are not related to the change of rest mass. In the case of no-dependence of the extra coordinate 
(\ref{new def for force per unit mass contr}) and (\ref{new def for force per unit mass cov}) become identical to (\ref{Classical force contr per unit mass}) and (\ref{Classical force cov per unit mass}) with $\partial g_{\alpha \beta}/\partial y = 0$. Consequently, there is complete consistency between the new expressions for the extra force and previous results in the literature.

The conclusion from the above is that the force from an extra non-compactified dimension does not necessarily contradict physics in $4D$. We propose our equations (\ref{new def for force per unit mass contr}) and (\ref{new def for force per unit mass cov}), instead of the abnormal force (\ref{Classical force contr per unit mass}) and (\ref{Classical force cov per unit mass}),  as the correct  expressions for the force from  a non-compactified extra dimension.

\subsection{The new definition at work}
Let us now consider the scenario discussed in section $3.3.3$. In this case $f^{\sigma}_{\Phi \perp} = 0$, $f^{\sigma}_{g \perp} = 0$, and because $m_{0}$ is constant we get $f^{\sigma}_{\parallel} = 0$. Which is what we expected from a physical point of view, instead of the unphysical prediction given by (\ref{fifth force contr}) and/or (\ref{fifth force cov}).

For the solution in section $3.3.4$, again $f^{\sigma}_{\Phi \perp} = 0$, $f^{\sigma}_{g \perp} = 0$, but now $f^{\mu}_{\parallel} \neq 0$, viz.,
\begin{equation}
\frac{f^{\mu}}{m_{0}} = \frac{(1 - 2\alpha)}{\alpha t}u^{\mu}.
\end{equation}
We note that the variation of rest mass $m_{0}$ provided here (and in section $3$) occurs on cosmic timescales. Therefore, this force would be very small as to be observed in a direct experiment. Our aim here has been  to illustrate what we can get from the theory rather than to formulate explicit experimental suggestions. The application of the formalism discussed here to other five-dimensional scenarios may lead to concrete observationally or experimentally testable predictions.

\section{Summary and conclusions}

In this paper we have provided a comprehensive discussion of how an observer in $4D$ perceives the five-dimensional geodesic motion of test particles. We discussed three aspects of the dynamics in $4D$; (i) the effective mass $m_{0}$, (ii) the trajectories, and (iii) the extra non-gravitational forces observed in $4D$.

Conventionally this topic is examined exclusively on the base of the splitting of the five-dimensional geodesic equation. Here we have used the geodesic equation but introduced a new element in the discussion which is the Hamilton-Jacobi (HJ) formalism. 

Since the mass of the particle does not appear anywhere in the geodesic equation, in section $2$ we used the HJ formalism to provide a practical equation, namely (\ref{HJ 4D}), for the computation of mass in $4D$. In that equation we use the action $S$ which is the solution of the five-dimensional HJ equation. Therefore, it transports  from $5D$ to $4D$ details of the extra dimension and  information of  the higher-dimensional motion. Consequently, the mass calculated in $4D$ from (\ref{HJ 4D}) depends on the signature of the extra dimension, the motion in the bulk and on whether the particle in $5D$ is massive or massless. Next we used the fourth component of the geodesic equation to obtain an explicit expression for the variation of rest mass along the particle's worldline, as observed in $4D$. 

The advantage of our formulation is that the mass and its variation are independent of the coordinates used and free of other ambiguities associated with the details of the dimensional reduction of the  geodesic, which are clearly not unique \cite{Seahra4}. Thus, we can claim here that variable rest mass $m_{0}$ is {\em not} an artifact of a poor choice of coordinates or parameter used in the geodesic description, as discussed in \cite{Seahra4}, but it is a four-dimensional manifestation of the extra dimension.

In section $3.1$ we discussed the trajectories in $5D$ confined to $y = const.$ hypersurfaces. It is clear that, in general, non-gravitational forces acting in the bulk, orthogonal to spacetime, are needed in order to keep the particles moving on such hypersurfaces. But  the  introduction of force fields, other than gravity, living in the bulk seems to be  in opposition to the initial spirit of the theory. On the other hand, given a metric in $5D$ we do not know what is the  metric frame that appropriately represents our four-dimensional spacetime. Therefore we conjecture that the choice of the metric frame on which the $4D$ theory lives should be limited by the physical condition of confinement, without introducing non-gravitational forces. 

In section $3.2$ we discussed the general, not confined, motion in $5D$. With the purpose of illustration, in $3.3$ we examined the motion of massive and massless test particles in the bulk cosmological metric (\ref{Ponce de Leon solution}). Our analysis clearly illustrates the variety of physical scenarios and the simplicity of the HJ method for the calculation of the corresponding masses and trajectories in $4D$. The explicit scenarios considered in $3.3.3$ and $3.3.4$ present the property that, although the motion is not assumed to be confined, it becomes confined to $4D$ spacetime asymptotically in time. This is an intriguing feature because the $5D$ metric (\ref{Ponce de Leon solution}) embeds the flat FRW models. This raises the  question of whether this feature is not a common one among higher-dimensional embeddings of our $4D$ world. If this were so, the confinement of matter to spacetime would come out automatically from the physics, without imposing extra conditions (or introducing non-gravitational forces).  The application of the formalism to other embeddings should help us to clarify this question.  

A test particle moving geodesically in the five-dimensional manifold is perceived in $4D$ to be moving under the influence of an extra force. In section $4.1$ we demonstrated, by means of an explicit example, that the quantity  calculated from (\ref{Definition of force in 4D contravariant}) and/or (\ref{definition of force in 4D covariant}) contradicts basic physics in $4D$. In particular, $F^{\mu}_{\parallel}$ is not produced by the variation of mass, because in $4.1$ the mass observed in $4D$ is constant. Then in $4.2$, we showed that this contradictory behavior is not limited to the scenario in $4.1$, but it is a generic one. 

Thus, we claim that (\ref{Definition of force in 4D contravariant}) and (\ref{definition of force in 4D covariant}) are {\em not} applicable for the case of large extra dimensions. They lead to wrong results by two separate reasons. First, in this case they do not represent the components of a four-vector. Secondly, the mass and/or its variation are not appropriately implemented, as evidence (\ref{fifth force contr}) and (\ref{fifth force cov}). This is not surprising because in previous work concerning this force there is no explicit computation of the rest mass or its variation along the particle's worldline.

In section $4.3$ we constructed the correct definition for this force as measured in $4D$. We have seen that although $u^{\mu}Du_{\mu} \neq u_{\mu}Du^{\mu} \neq 0$ and $Dg_{\mu\nu} \neq 0$, the equations are consistent with the usual properties  $u^{\mu}D^{(4)}u_{\mu} =  u_{\mu}D^{(4)}u^{\mu} = 0$ and $D^{(4)}g_{\mu\nu} = 0$, provided the absolute derivative in $4D$ is defined as in (\ref{operator D4}). This definition allowed us to get rid of the parallel term $u^{\mu}(u^{\alpha}u^{\beta}\partial g_{\alpha \beta}/\partial y)$, which was initially associated to the fifth force, but is not related to the change of mass. The final expressions for the extra force are provided by (\ref{new def for force per unit mass contr}), (\ref{new def for force per unit mass cov}). They satisfy appropriate physical conditions and extend previous work in the literature  \cite{JPdeLforce}, \cite{Seahra3} and \cite{Eq. of Motion}.
   
In our approach, in order to calculate $f^{\mu}_{\parallel}$ we first find the five-dimensional action from (\ref{HJ 5D}), then from (\ref{HJ 4D}) we obtain $m_{0}$ as a function of coordinates. Next we calculate $dm_{0}/ds$ and evaluate the final expression along the trajectory. The result depends on the details of the motion in $5D$. An example of this procedure was  shown in section $4.4$.

We notice that in our discussion the underlying physics motivating the introduction of  a large extra dimension was nowhere used. Neither, the physical meaning of the extra coordinate. Therefore, our  results are applicable to brane-world models, STM, and other $5D$ theories with a large extra dimension. In STM and in the thick brane scenario the $5D$ manifold is smooth everywhere and there are no defects. 

In the RS2 brane-world scenario \cite{RS2} our universe is a singular $4D$ hypersurface and the derivatives $\partial g_{\mu \nu}/\partial y$ are discontinuous, and change sign, through the brane. However, the discontinuity is not observed \cite{Seahra3} and effective $4D$ equations can be obtained by taking mean values and applying Israel's junction conditions through the brane \cite{Youm2}. 

In summary, our work provides a unified approach for the discussion of mass, trajectories and forces acting on test particles. Important new results here are (i) the invariant definition of mass  and the equation for the variation of mass along the particle's worldline in $4D$, (ii) the new definition for the extra forces which does not contradict physics in $4D$, and (iii) the characterization of motion in $4D$ and $5D$ without having to deal with the ambiguities associated with the splitting of the geodesic equation.  The latter allowed us to  confirm and give an improved description of many results previously obtained in the literature.

\end{document}